\documentclass[11pt]{article}
\usepackage{amsmath} 
\usepackage{amssymb} 
\usepackage{graphicx}
\usepackage{url}
\usepackage[round]{natbib}
\usepackage{setspace}
 \usepackage[small]{caption}
\begin{document}
\author{David R.B. Stockwell\thanks{http://landshape.org, Email: davids99us@gmail.com}~ and Anthony Cox} 
\date{\today}
\title{Structural break models of climatic regime-shifts: claims and forecasts} 
\maketitle

\section{Abstract}

A Chow test for structural breaks in the surface temperature series is used to investigate two common claims about global warming. Quirk (2009) proposed that the increase in Australian temperature from 1910 to the present was largely confined to a regime-shift in the Pacific Decadal Oscillation (PDO) between 1976 and 1979. The test finds a step change in both Australian and global temperature trends in 1978 (HadCRU3GL), and in Australian rainfall in 1982 with flat temperatures before and after. Easterling \& Wehner (2009) claimed that singling out the apparent flatness in global temperature since 1997 is 'cherry picking' to reinforce an arbitrary point of view. On the contrary, we find evidence for a significant change in the temperature series around 1997, corroborated with evidence of a coincident oceanographic regime-shift. We use the trends between these significant change points to generate a forecast of future global temperature under specific assumptions. 

\section{Introduction}

Climatic effects of fluctuations in oceanic regimes are generally studied by decomposing rainfall and temperature into periodic components: e.g. singular spectrum analysis (SSA) \citep{Ghil:1991th}, and variations on principle components analysis (PCA) \citep{Parker:2007nx}.  Such approaches can capture such effects as the influence of short period phenomena like El Nin\~{o} on Australia \citep{Mantua:1997kl, Guilderson:1998nx}, and the potential for longer term phenomena such as the Pacific Decadal Oscillation (PDO) to 'offset' putative increases in global temperature \citep{Keenlyside:2008dq}.  Methods designed for finding and testing structural breaks address more infrequent regime-shifts.  An F-statistic known as the Chow test \citep{Chow:1960hc} based on the reduction in the residual sum of squares through adoption of a structural break, relative to an unbroken simple linear regression, is a straightforward approach to modeling regime-shifts  with structural breaks.  

Two claims are evaluated here. In the first,  \citet{Quirk:2009sf} proposed a model of Australian temperature with a regime-shift (herein Q09): slightly increasing to 1976, rapidly increasing to 1979 (the shift), and slowly increasing therafter.  The increase in Australian temperature of around 0.9$^\circ$C from the start of the readily available records in 1910 \citep{BoM} is conventionally modeled as a linear trend and, despite the absence of clear evidence, often attributed to increasing concentrations of greenhouse gases (GHGs) e.g. \citep{CSIRO:la}.  \citet{Quirk:2009sf} proposes a different model, without testing its significance.  Regarding only the trends of 0.034$^\circ$C and 0.05$^\circ$C per decade in the sections before and after the shift substantially lowers the underlying rate of warming potentially attributable to anthropogenic global warming (AGW).    

\citet{Hartmann:2005yq} noted for the climate of Alaska,  that recognition of an abrupt temperature change in climate 1976 profoundly changed the underlying temperature trends from strongly positive, to slightly negative, contradicting a view of increasing greenhouse gases (GHGs) gradually increasing temperatures.  The event dubbed the Great Pacific Climate Shift \citep{Kerr:1992fy}, is clearly seen in a change in the average polarity of the Pacific Decadal Oscillation (PDO) from a cool to a warm phase \citep{Hartmann:2005yq}. 

The second example is a claim of lack of statistical significance.  Regarding the declining temperature since the El Nin\~{o} event in 1998, \citet{Easterling:2009wd} state that using 1998 as a starting point is 'cherry-picking' to justify a point of view and therefore, statements such as 'global warming stopped in 1997' have no basis.  Structural change methods are used to determine if the period is a significant change point. 

From a broader perspective, econometrics research into regime-shifting models indicates that detecting breaks in time series is important and ignoring them can be very costly \citep{Pesaran:2004dp}.  The cost in climate forecasting would be inaccurate forecasts.  That is, while a forecast based on a linear model would indicate steadily changing global temperatures, forecasts based on shifts would reflect the moves to relatively static mean values.  The choice of underlying model may also impact estimates of the magnitude of climate change, as the \citet{Quirk:2009sf} attribution of a trend of either 0.04$^\circ$C or 0.09$^\circ$C shows. Further, oceanographic changes have been postulated as the main causes of changes in global temperature in the last 30 years \citep{gray:2007}, including the recent flat to slightly declining trend in temperatures since 1997.

A larger relevance of structural change models to global temperature data is suggested by the presence of long term persistence (LTP) in complex natural systems \citep{Kout:2008tg, Halley:2009ai}.  It is known that step changes in the mean at multiple scales can reproduce LTP behaviour \citep{Kout:02, Stockwell:2006rf}.  Change points can be found in the mean, or standard deviation \citep{Bai:1994oq}. 

There are well-known difficulties in forecasting with unstable structural breaks models \citep{Hamilton:1989sp, Hamilton:2007ij}. In particular it is known that a least-squares estimator of change points is not reliable for nonstationary $I(d)$ data with $0.5 < d < 1.5$ \citep{Hsu:2007eu}.   But when there is a change, the least squares estimator is known to be reliable on stationary $I(d)$ data (i.e., -0.5 < d < 0.5).   Here $d$ is the fractional differencing parameter as introduced by \citet{Granger:1980qe}.  We also have some reservations about the data being of adequate length, as models induced from the temperature record can be contingent on subsequent data \citep{Stockwell:2009kx}.  Given these uncertainties, it is important to develop models rigorously, with a transparent development sequence satisfying both significance and optimization of the model space.

\section{Method}

We attempt to verify the optimality and significance of the breakpoints, by reference to both climate data and regime-shifts.  Australia-wide annual average rainfall and temperature anomalies were downloaded from the BoM data site \citep{BoM}.  Temperature is available as an annual average in anomaly $^\circ$C from 1910 to 2008 while rainfall in anomaly millimeters covers 1900 to 2008.   \citet{Quirk:2009sf} only shows temperature data from 1950 to 2003.  We also examine the global monthly temperature anomalies from the Hadley Center, HadCRU3GL, for confirmation of the presence of breakpoints at the larger scale \citep{Rayner:2006kx}.  Comprehensive evaluation of the full range of available global datasets was beyond the scope of this study.  

The approach to developing a structural break model was to calculate the F statistic (Chow test statistic) for potential breaks at all potential change points using the R package $strucchange$ \citep{Zeileis:2002cr}. In this test, the error sum of squares (ESS) of a linear regression model is compared with the residual sum of squares (RSS) from a model composed of linear sections before and after each potential change point.  If $n$ is the number of observations and $k$ the number of regressors in the model, the F statistic is:

$F = (RSS-ESS)/ESS * (n-2*k)$

Potential breakpoints are indicated by peaks in the value of the F statistic (dashed vertical lines in figures).  The red line indicates the height above which a series with a break at that location would be significantly better (at the 95\% confidence limit) than a single linear regression.

\begin{table}[ht]
\begin{center}
\caption{Fit of models as described in text and shown in Figures 1a,b and 2a,b.  R2 is the regression correlation coefficient, R2b is  the correlation coefficient with one break, Pb is significance of improvement of the break, over the linear model, P1 and P2 and the significances of the slopes of the first and second segments, respectively.}
\begin{tabular}{rrrrrrr}
  \hline
 & R2 & Breakdate & R2b & Pb & P1 & P2 \\ 
  \hline
Aus. Temperature & 0.384 & 1978.00 & 0.482 & 0.000 & 0.151 & 0.139 \\ 
  Aus. Rainfall & 0.041 & 1982.00 & 0.164 & 0.017 & 0.877 & 0.026 \\ 
  Global Temp. 1910-2008 & 0.721 & 1978.00 & 0.785 & 0.000 & 0.000 & 0.000 \\ 
  Global Temp. 1976-2008 & 0.613 & 1997.08 & 0.672 & 0.000 & 0.000 & 0.409 \\ 
   \hline
\end{tabular}
\label{tab1}
\end{center}
\end{table}

The relative fit of the models is indicated with a regression coefficient $R^2$, as listed in Table~\ref{tab1}.  In addition, $p$ values indicate the significance of model improvements (Pb), and the non-zero slope parameters (P1 and P2).

While the $strucchange$ software also implements a search for multiple breakpoints, we adopt a conservative approach without undue complexity, mindful of the problematic nature of simultaneous validation of multiple structural break models \citep{Zeileis:2004rr}. The optimal position of single breaks is clearly indicated with a plot of the F statistic aligned with the x-axis of a plot of the time series.

 \section{Results}

As a preliminary check, the estimates of the fractional differencing parameter $d$ on the Australian temperature and rainfall data were estimated at 0.31 and 0.13 respectively (using R package $fracdiff$).   Therefore, the Chow test is expected to be reliable.  The estimate of $d$ on HadCRU3GL global temperature series is 0.50, so results must be interpreted more cautiously.

\begin{figure}
\makebox[\textwidth][l] {
\includegraphics[width=1\textwidth]{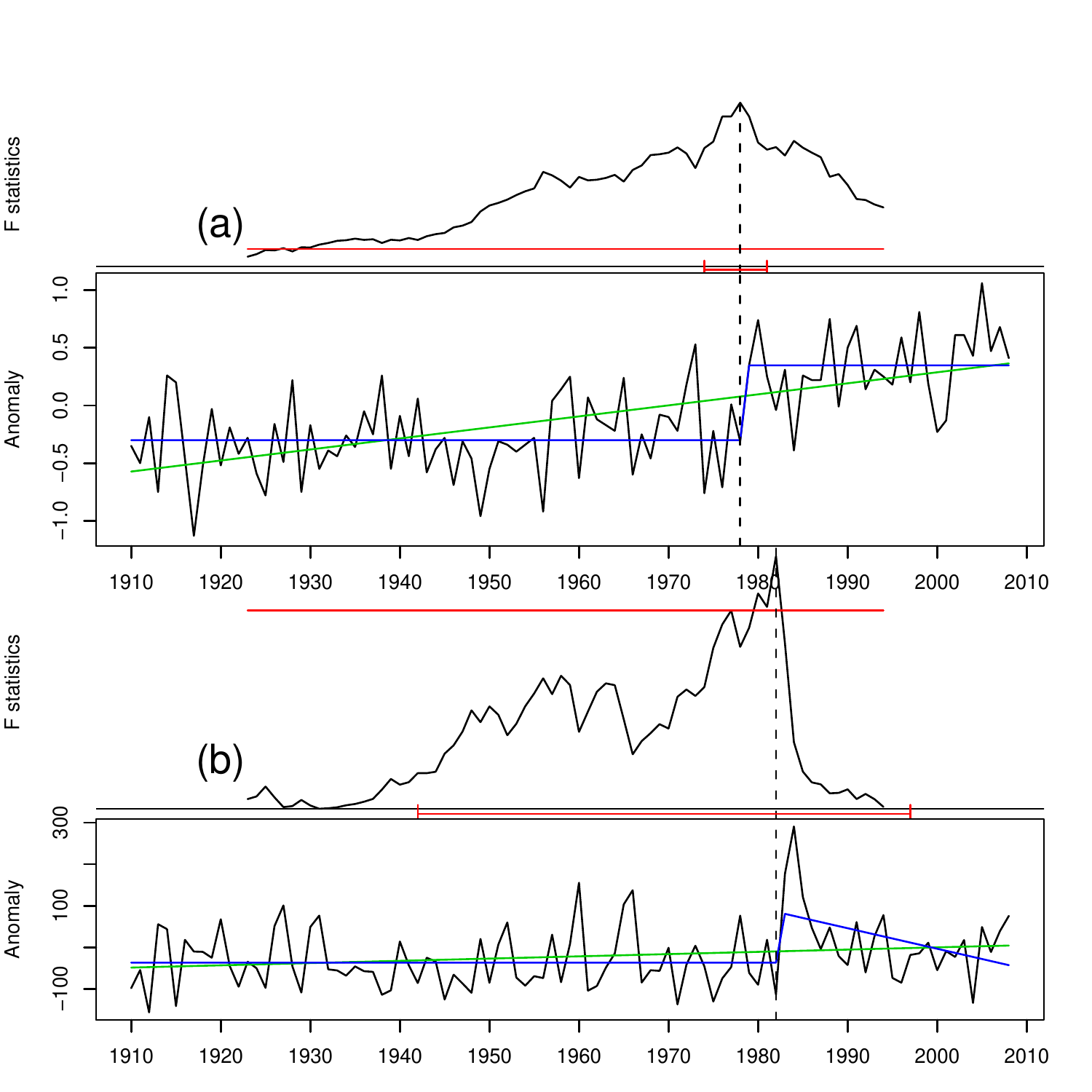}
}
\caption{Breakpoints in Australian temperature (a) and rainfall (b).  The upper section of each figure is the F statistic from the Chow test for breaks with optimal breaks (vertical, dashed), level of significance (red line) and confidence interval (red bracket). The Australian temperature (below), with the OLS regression model (green) and the optimal break model (blue).}
\label{fig1}
\end{figure}

The F statistics, breakpoints, confidence intervals, the simple regression and segmented models for annual Australian temperature from 1910 to 2008, are shown on Figure~\ref{fig1}a.  The peak in the F statistic above the red line indicates that 1978 is the optimal location for a break, indicated with a dashed vertical line.  The 95\% confidence interval spans 1974 to 1981, indicated with a red bracket on the x-axis of the F statistic.  The structural break model is a significant  ($p<0.001$) improvement in the fit of the model as shown by the increase in $R^2$ from 0.38 to 0.48.   The slope of the two segments is not significant ($p=0.15$ and $p=0.14$).  Because of the significance of the break, and the non-significence of slopes in the segments, a model with a single step, or structural change model is justified (Figure~\ref{fig1}a blue).  

Regarding the precipitation series, a single break-date at 1982 (Figure~\ref{fig1}b) significantly improves the fit ($p=0.02$) as indicated by the increase in $R^2$ from 0.041 to 0.164 (Table~\ref{tab1}).  The trend in the segment before the break is not significant ($p=0.88$), but a decline in precipitation after the break is significant ($p=0.03$), most probably due to the anomalously high rainfall in the years after 1978.

\begin{figure}
\makebox[\textwidth][l] {
\includegraphics[width=1\textwidth]{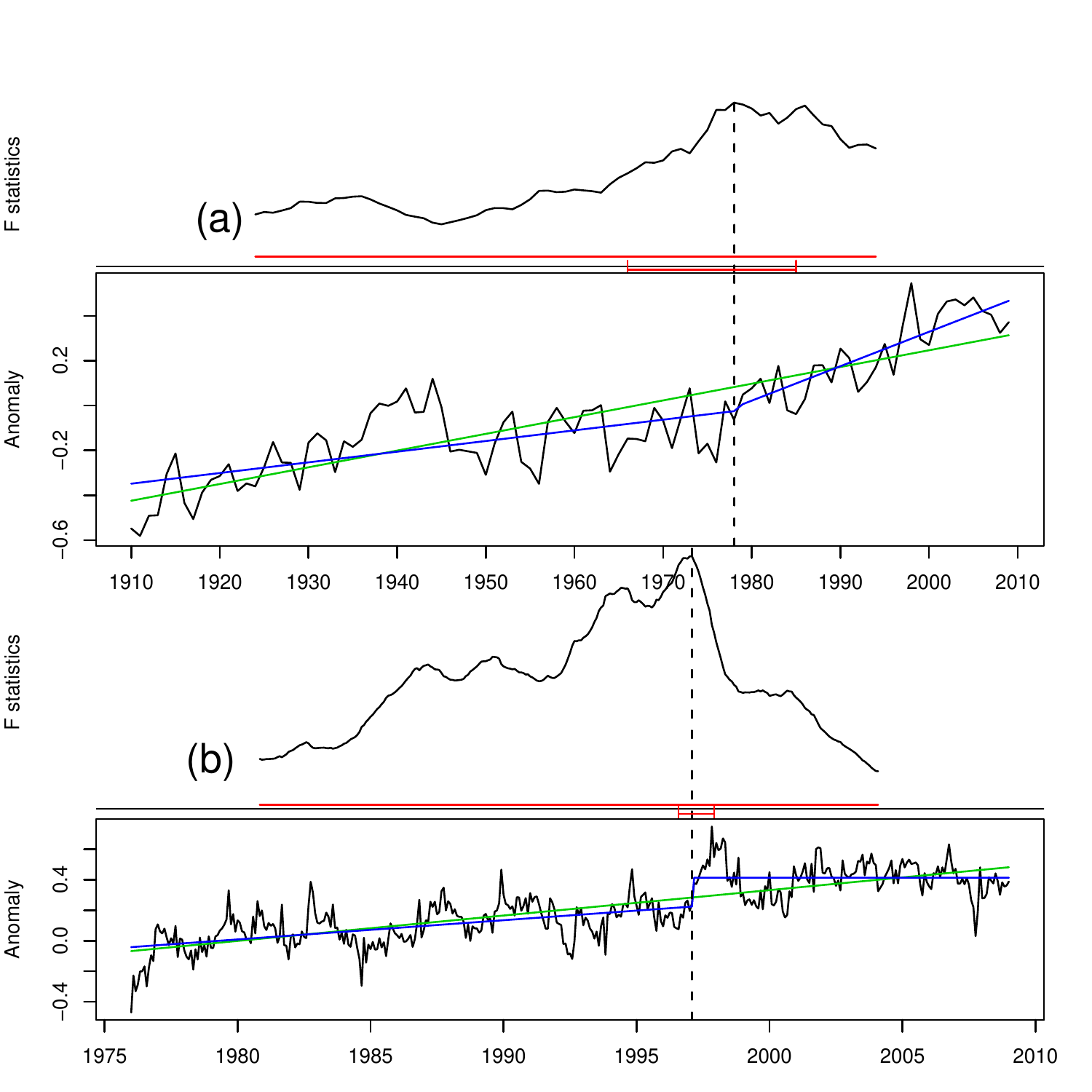}
}
\caption{Breakpoints in sea surface monthly temperature anomalies (HadCRU3GL) (a)1910 to present and (b) 1976 to present.}
\label{fig2}
\end{figure}

A broad peak in the F statistic for annual global surface temperature from 1910 at 1978 (Figure~\ref{fig2}a) ($p<0.001$) improves the fit of the model from an $R^2$ of 0.56 to 0.68.  The 95\% confidence interval spans 1966 to 1985, indicated with a red bracket below.  When the monthly global surface temperature series is truncated to the break-date of 1978, the F statistics indicate a strong break-date in 1997(8) (Figure~\ref{fig2}b), $R^2$ increasing from 0.61 to 0.67) with a rising trend prior to 1997, and a flat trend thereafter ($p=0.41$).  The 95\% CI spans 1996(8) to 1997(12).

\section{Discussion}

These results confirm a break-date around 1978, as suggested by the general form of Q09  \citep{Quirk:2009sf}, but differs in respect to one, not two breaks, with non-significant trends outside of the optimal break-date.  This is known as a change point model, characterized by abrupt changes in the mean value of the underlying dynamical system, rather than a smoothly increasing or decreasing trend.  The confidence in 1978 as a break-date is further strengthened by the results for global temperatures since 1910, indicating the series could be described as gradually increasing to 1978 ($0.05\pm0.015^\circ$C per decade), with a steeper trend thereafter ($0.15\pm0.04^\circ$C per decade).

The Chow test since 1978 finds another significant break-date in 1997, delineating an increasing trend up to 1997 ($0.13\pm0.02 ^\circ$C per decade) and non-significant trend thereafter ($-0.02\pm0.05^\circ$C per decade).  Contrary to claims in \citet{Easterling:2009wd} that the 10 year trend since 1998 is arbitrary, structural change methods indicate that 1997 was a statistically defensible beginning of a new, and apparently stable regime.  

The significance of the dates around 1978 and 1997 to climatic regime-shifts is not in dispute, as they are associated with a range of oceanic, atmospheric and climatic events, whereby thermocline depth anomalies associated with PDO phase shift and ENSO were transmitted globally via ocean currents, winds, Rossby and Kelvin waves \citep{Guilderson:1998nx, McPhaden:2004fv, Wainwright:2008eu}.  Even though the pattern of response would vary in different parts of the globe, a step-change in temperature remarkably similar to Australia also occurred in Alaska \citep{Hartmann:2005yq}.

Given the assumption that AGW is not a cause of regime-shifts, an assumption that would require further verification, these results suggest a considerably stronger claim for Australian climate, i.e. because the slope of the two segments is not significant (p=0.15 and p=0.14), any trend due to increasing $CO_2$ is statistically insignificant.

Decomposition of global temperature into regime-shifts and AGW is more ambiguous, but it would be argued, similar to Q09, that a large proportion of the increase in temperature increase between 1978 and 1997 was associated with the propagation of a major regime-shift, with low rates of AGW in the segments before and after.  Supporting this view are such oceanographic observations as the large reduction in the rate of Pacific equatorial upwelling around 1976 and its resumption in 1998, a reduction of westward volume transport between Australia and Indonesia of 23\% from 1976-7 and a general elevation of sea temperature by 1-2$^\circ$C \citep{Wainwright:2008eu}. Following two decades of weaker flow coinciding with rising global temperatures from the break-date of 1978, convergence of cold interior ocean pycnocline water towards the equator increased from 13.4 to 24.1$m^3s^{-1}$, and Pacific sea surface temperatures cooled \citep{McPhaden:2004fv}.  \citet{Cai:2007lq} asserts a role for strengthening of the global conveyor in global temperatures, by intensifying the rate of heat transfer out of the off-equatorial region and into the subtropics. The relative contribution and interaction of various ocean phases: IPO, PDO and ENSO, the duration of lags, and precipitation effects are more uncertain \citep{Power:1999fv}.

One can still argue for human influence on regime-shifts.  \citet{Cai:2007lq} attributes changes in the global conveyor to anthropogenic aerosols, and \citet{Vecchi:2006dq} attributes a weakening of the Walker circulation to anthropogenic forcing. Neither attribute abrupt regional climate changes to non-natural causes.  There is also the view that stable temperatures since 1997 may be a result of an 'offset' or masking of AGW effect by natural variation \citep{Keenlyside:2008dq}.  For consistency, this view must also entertain the possibility of putative AGW being amplified by natural variability in the previous regime from 1978 to 1997. 
     
Assuming a regime-shift from 1978-98 reduces the estimates of the underlying rate of AGW warming from around $0.14^\circ$C to  $0.05^\circ$C per decade.  An increase of $0.5^\circ$C by 2100 is consistent with low-end empirical estimates of climate sensitivity, such as \citet{Spencer:2008kx} at $0.6^\circ$C for $2XCO_2$, but considerably lower than the IPCC projections for the most common AGW scenarios \citep{IPCC:2007ya}.  Our model is consistent in period and timing to the two component model proposed by \citet{Akasofu:2009kx}, with a contribution from a natural multidecadal oscillation of $0.15^\circ$C per decade between 1975 and 2000, and a  steady contribution of $0.05^\circ$C per decade from the upswing of a multi-centennial oscillation, such as the natural cycle of glacial formation and regeneration called the Great Season Climatic Oscillation \citep{boucenna-2008}.  In our case, the model was developed from statistically justified, empirical analysis of temperature data.  

\begin{figure}
\makebox[\textwidth][l] {
\includegraphics[width=1\textwidth]{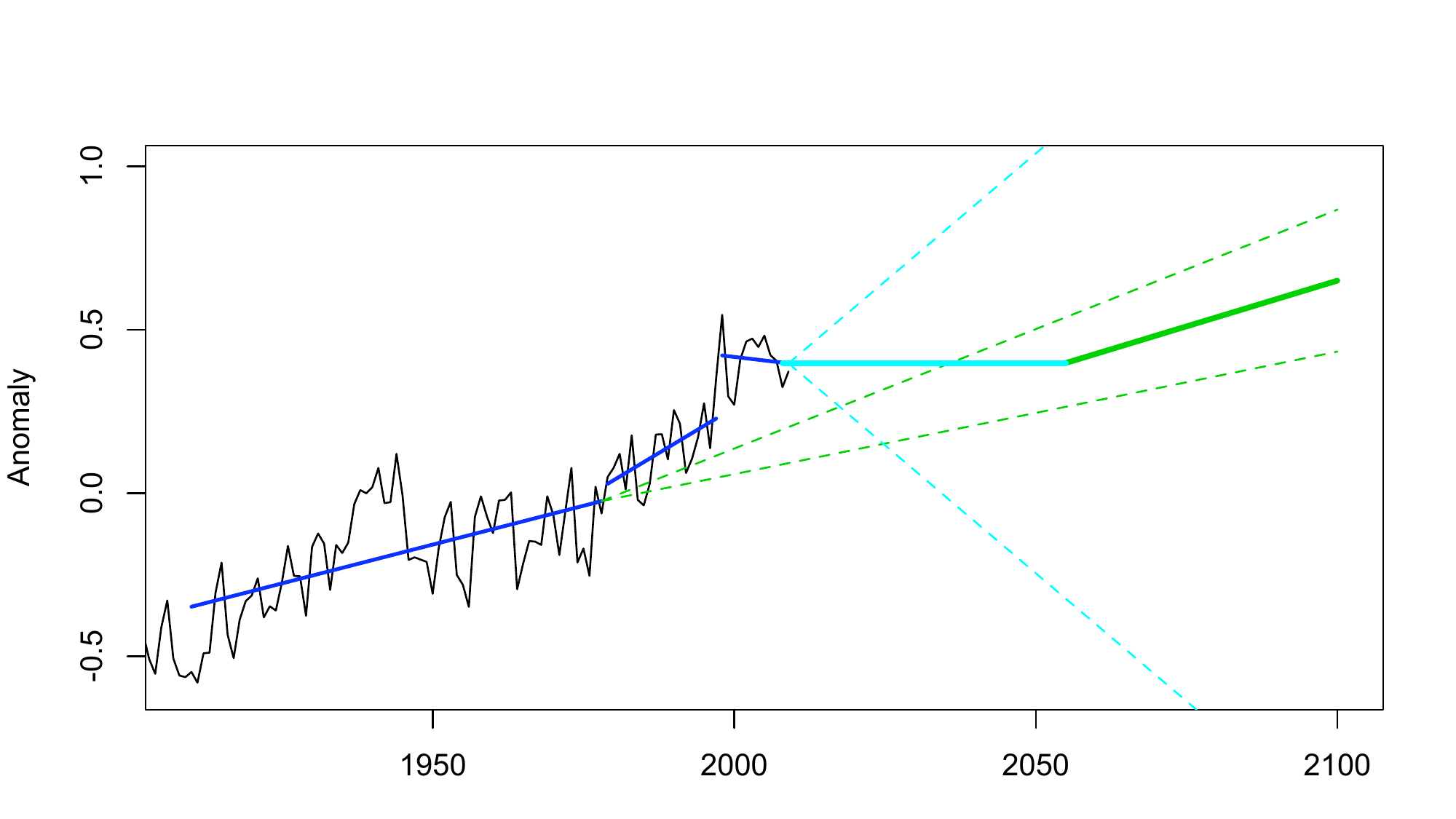}
}
\caption{Prediction of global temperature to 2100, by projecting the trends of segments delineated by significant regime-shifts.   The flat trend in the temperature of the current climate-regime (cyan) breaks upwards around 2050 on meeting the (presumed) underlying AGW warming (green), and increases slightly to about $0.2^\circ$C above present levels by 2100.  The 95\% CI for the trend uncertainty is dashed. }
\label{fig3}
\end{figure}

Figure~\ref{fig3} illustrates the prediction for temperatures to 2100 following from our structural break model, the assumptions of continuous underlying warming, regime-shift from 1978 to 1997, and no additional major regime-shift.  The projections formed by the trend to 1978 (presumed AGW warming, green) and the trend in the current regime (AGW offset by regime-shift, cyan) predicts constant temperatures for fifty years to around 2050, similar to the period of flat temperatures from 1930-80, then increasing to about $0.2^\circ$C above present by 2100.   It must be kept in mind that this extrapolation is based on greatly simplified assumptions regarding regime-shifts and trends, and does not incorporate many of the complexities and natural forcing factors operating to produce climate change in the real world \citep{marsh-2009}.

These results justify further development of more complex regime-shift models of temperature and rainfall for forecasting purposes, including attempting to decompose global climate into temporally and spatially differentiated regime-shift models.

\section{Acknowledgements}

We gratefully acknowledge the comments of John McLean and David Evans on earlier versions of the manuscript.  

\bibliographystyle{article} 
\bibliography{../../mylibrary}

\end{document}